\documentclass[nofootinbib, twocolumn,preprintnumbers,amsmath,amssymb, superscriptaddress]{revtex4-1}
\usepackage{amsmath, amsfonts, amsthm, amssymb, graphicx}
\usepackage[utf8]{inputenc}
\usepackage{ulem}
\usepackage{bm}
\usepackage{tabularx}
\usepackage{float}
\usepackage{placeins}

\newcolumntype{Y}{>{\centering\arraybackslash}X}

\renewcommand{\vec}[1]{\bm{#1} }

\usepackage[dvipsnames]{xcolor}

\def\changes{\textcolor{black}}
\def\changesNew{\textcolor{black}}

\begin{document}

\title{Nonclassicality of axion-like dark matter through gravitational self-interactions}

\author{Michael Kopp}
\email{michael.kopp@su.se}
\affiliation{Department of Physics, Stockholm University, SE-106 91 Stockholm, Sweden}
\affiliation{Nordita,
KTH Royal Institute of Technology and Stockholm University,
Hannes Alfv\'ens v\"ag 12, SE-106 91 Stockholm, Sweden}

\author{Vasileios Fragkos}
\email{vasileios.fragkos@fysik.su.se}
\affiliation{Department of Physics, Stockholm University, SE-106 91 Stockholm, Sweden}

\author{Igor Pikovski}
\email{igor.pikovski@fysik.su.se}
\affiliation{Department of Physics, Stockholm University, SE-106 91 Stockholm, Sweden}
\affiliation{Department of Physics, Stevens Institute of Technology, Hoboken, NJ 07030, USA }

\date{\today}

\begin{abstract}

Axion-like particles (ALPs) are promising dark matter candidates. They are typically described by a classical field, motivated by large phase space occupation numbers. Here we show that such a description is accompanied by a quantum effect: squeezing due to gravitational self-interactions. For a typical QCD axion today, the onset of squeezing is reached on $\mathrm{\mu s}$-scales and grows over millennia. Thus within the usual models based on the classical Schr\"odinger-Poisson equation, a type of Gross-Pitaevskii equation, any viable ALP is nonclassical.
\changes{We also show that squeezing may be relevant on the scales of other self-gravitating systems such as galactic haloes, or solitonic cores.} Conversely, our results highlight the incompleteness and limitations of the classical single field description of ALPs.
  \end{abstract}
\maketitle

\section{Introduction}

The search for dark matter is one of the main challenges of modern physics. A viable and popular dark matter candidate is a light (mass $m \ll 1\,\rm{eV}$) self-gravitating quantum scalar field, such as the QCD axion or any other axion-like particle (ALP). Extensive experimental searches  focus on its distinct wavelike signatures \cite{SchiveChiuehBroadhurst2014, PozoBroadhurstdeMartinoEtal2020, ArvanitakiDimopoulosGalanisEtal2020}, and its theoretically expected coupling to matter \cite{Marsh2016, Sikivie2021, BackesPalkenKenanyEtal2021}.
This type of ALP dark matter, also including ultra-light or ``fuzzy dark matter,'' is usually described by a classical scalar field rather than particles
\cite{AbbottSikivie1983, PreskillWiseWilczek1983, DineFischler1983, WidrowKaiser1993, Peebles2000, HuBarkanaGruzinov2000, GuthHertzbergPrescod-Weinstein2015, Hertzberg2016, HuiOstrikerTremaineEtal2017, Hui2021}.
The classical field description is motivated by the  de Broglie wave length $\lambdabar_{\rm deB}=\hbar/(m v_{\rm rms})$ being significantly larger than the inter-particle distance $d$, and thus the mean phase space occupation number $(\lambdabar_{\rm deB}/d )^3$
is large.
If a misalignment mechanism produces the ALP then initially it would be in a coherent state \cite{AbbottSikivie1983, PreskillWiseWilczek1983}, and thus a classical field.

An open question is whether dark matter can exhibit observable quantum features.
The difference between quantum and classical dynamics can be subtle, and there are several non-classicality measures in quantum information science \cite{goldberg2020extremal}.
One such non-classicality witness is the inability to represent the system in terms of classical mixtures of coherent states. The most prominent quantum effect of this form is squeezing \cite{drummond2013quantum}. In the context of quantum optics, squeezing of bosonic systems is a well established benchmark for non-classicality that has been observed in a range of different systems \cite{TanasKielich1983, BreitenbachSchiller1997, Dodonov2002, wollman2015quantum}, with many applications in quantum-enhanced metrology \cite{GrossZiboldNicklasEtal2010} and quantum information processing \cite{zhong2020quantum, hosten2016measurement}. \changesNew{The electromagnetic case is also instructive to understand the difference between the classical and quantum states of the ALP. The starting point of quantum optics and our understanding of the difference between classical and quantum coherence was the Hanburry-Brown-Twiss experiment \cite{brown1954lxxiv}. It focused on the correlation between intensities from a beam of light that is received at two spatially separated detectors. While the observed correlations can be described classically, Glauber \cite{glauber1963coherent} and Sudrashan \cite{sudarshan1963equivalence} used it to develop a quantum description of light that showed how one could distinguish classical from quantum effects:  photon number correlations can exhibit classical or quantum features -- the latter typically referred to as anti-bunching. Squeezed light is a particular quantum state that may exhibit such quantum signatures that can be verified experimentally. As we will show in this work, ALPs may well be in such a state.}

In this paper we consider the dynamics of a self-gravitating quantum field and show that it undergoes rapid self-squeezing due to self-interactions, a quantum feature that cannot be captured by a classical field description.
Our results hold in the standard framework used for describing a classical scalar field, the Gross-Pitaevskii equation (GPE) in the form of the Schr\"odinger-Poisson equation.
The squeezing happens on $\mu s$-timescales and is continuously reproduced. We show formation of squeezing in three different physical scenarios: \changes{ on cosmological scales, galactic scales, and in solitonic cores that form in a galactic centers. These scenarios have the common feature that they are approximately self-gravitating.  On even smaller scales, such as haloscopes (ground based experiments searching for axions) \cite{Sikivie1983} we find that squeezing becomes negligible even if the galactic halo undergoes strong squeezing}.
A QCD axion, for example, reaches $9 \mathrm{dB}$ squeezing -- reduction of the vacuum noise for one quadrature (and simultaneous increase for the opposite quadrature) by an order of magnitude -- after less than $100 \mu s$ for the cosmological volume \changes{down to the solitonic core volume}.
Thus, under standard assumptions about the dynamical description of the ALP, we arrive at the result that an initial ``classical'' quantum state quickly evolves into a non-classical squeezed state. In contrast to previous works \cite{Hertzberg2016, DvaliZell2018}, here we find an example of quantum behavior that forms very rapidly. The results suggest that either ALPs exhibit non-negligible quantum features or that the usually adopted classical description by a single classical GP scalar field is incomplete.

\section{The ALP model}
We model the axion-like particle using a non-relativistic scalar quantum field $\hat \psi(\vec x, t)$. In a cosmological setting it is embedded in a homogeneously and isotropically expanding  Friedmann-Robertson-Walker (FRW) universe.
The resulting Heisenberg equation of motion of $\hat \psi(\vec x, t)$ is, see for instance\,\cite{KoppVattisSkordis2017},
\begin{align} \label{GPSPEHeisenberg}
    i \hbar \partial_t \hat \psi(\vec x, t) &= -\frac{\hbar^2}{2 m A(t)^2} \vec{\nabla}^2 \hat \psi(\vec x, t) + m \hat \Phi(\vec x, t) \hat \psi(\vec x, t) \\
    \vec{\nabla}^2 \hat \Phi(\vec x, t) &= \frac{4 \pi G m}{A(t)} \Big( \hat \psi^\dagger(\vec x, t) \hat\psi(\vec x, t) - \overline{\hat \psi^\dagger(\vec x, t) \hat\psi(\vec x, t)} \Big) \,, \notag
\end{align}
where $A(t)$ is the scale factor, which appears  due to our use of comoving coordinates $\vec{x}$. We remove the homogeneous mode, denoted by the overbar, from the source of the gravitational potential in the Poisson equation.
Our approach therefore does not quantise the homogeneous mode of the ALP. This allows us to avoid an IR divergence and conceptual issues of such a Newtonian quantum cosmology.
The homogeneous  mode obeys the Friedmann equation  $H(t)^2 = \frac{8 \pi G}{3} \rho_0\, A(t)^{-3}$. Here, $A(t)$ is the scale factor of the universe normalised to $A(t_0)=1$, where $t_0$ is present age of the universe. $H= \dot A(t)/A(t) $ is the Hubble expansion rate, and $\rho_0 = m n_0$ the present-day mean energy density of the ALP, with $n_0$ the present ALP mean number density and $m$ the mass of the ALP. We set the speed of light $c=1$.
For simplicity,  we neglect here other types of matter and the cosmological constant.\footnote{\changes{Inclusion of a cosmological constant is straightforward and would not affect our approach based on \eqref{GPSPEHeisenberg},  the order of magnitude of our results and our conclusions. On the other hand, inclusion of additional degrees of freedom will affect the quantum coherence, as discussed in section \ref{sec:discussion}.}}
To be consistent with the gravitational potential being entirely due to the ALP in \eqref{GPSPEHeisenberg} we focus on a purely ALP dominated background cosmology.
For our purposes, this is a sufficiently accurate description of the late universe.
While a cosmological constant could be trivially included (and will not change the order of magnitude of our results), inclusion of other types of  clustering matter would require a quantum mechanical modelling which is beyond the scope of this paper, but will likely play a crucial role in the quantum-classical transition of the ALP.

\subsection{Hartree ansatz}

Our approach to solve the Heisenberg equation  \eqref{GPSPEHeisenberg} is to impose the Hartree ansatz for the wave function of the ALP quantum field, and -- crucially -- the assumption that this ansatz remains dynamically valid, which guarantees the validity of the GPE \cite{ParkinsWalls1998, PitaevskiiStringari2003, AlonStreltsovCederbaum2007, ErdosSchleinYau2010}.
By doing so, we treat the ALP in a conservative way since the GPE remains valid, but at the same time obtain an analytically tractable nonlinear quantum description.
The large squeezing at very short timescales that we find here is precisely due to guaranteeing the validity of the GPE by assuming the Hartree ansatz. Thus our results show that squeezing should arise whenever this ansatz is justified.

The Hartree ansatz for the ALP quantum state implies that only a single mode with mode function $\psi(\vec x, t)$ and operator $\hat a(t)$ is relevant such that approximately
\begin{equation} \label{GPSPEHartree}
    \hat \psi(\vec x, t) = \frac{\psi(\vec x, t)}{\sqrt{N}} \hat a(t)\,,
\end{equation}
with all other modes $\widehat{\delta \psi}(\vec x, t)$ contributing to $\hat \psi(\vec x, t)$ neglected. \changes{This means that negligence of these other modes defines $\psi(\vec x, t)$ and  $\hat a(t)$ through \eqref{GPSPEHartree}.}
The normalisation $\sqrt{N}$ is for convenience and implies that the resulting GPE, for quantum states $| \Psi(t_i) \rangle$  containing $N$ (or approximately $N$) particles, will not depend on  $N$ in the limit $N \rightarrow \infty$.\footnote{The GPE field is normalised to $\int_V d^3\! x |\psi(\vec x,t)|^2 = N$, so that the spatial average is $\overline{|\psi|^2} = N/V \equiv n_0$.  Here $V$ is the comoving volume, assumed to be very large, roughly the size of the Hubble volume $H^{-3}_0$.}
Examples of such Hartree states are ``non-classical'' Fock states $| N \rangle= \frac{1}{\sqrt{N!}}(\hat a^\dagger(t_i) )^N |0\rangle$,
or ``classical'' coherent states   $| \alpha \rangle = e^{- N/2}\sum^\infty_{n=0} \frac{\alpha^n}{\sqrt{n!}} |n \rangle$,
 with $\alpha = \sqrt{N} $ so that  $\hat \psi(x,t_0) | \alpha \rangle = \psi(x,t_0) | \alpha \rangle$. The state $| \alpha \rangle$ describes a classical field configuration $\psi(x,t_0)$ with negligible quantum fluctuations $\mathrm{Var}(\hat{ \psi})^{1/2}/\psi = N^{-1/2}$.

\subsection{System of Kerr and Gross-Pitaevski equations}

\changes{Projecting the Heisenberg equation \eqref{GPSPEHeisenberg} onto the mode function,  that is  $ \frac{1}{\sqrt{N}\hbar}\int_V d^3\! x (\psi^*(\vec x,t) \,\mathrm{Eq.}(1))$,} we obtain the Kerr oscillator equation for $\hat a(t)$:
\begin{subequations} \label{GPSPKerrE}
\begin{align}
    i \partial_t  \hat a(t) &=  \omega(t)  \hat a(t) + 2\chi(t) \hat a^\dagger(t)  \hat a(t)  \hat a(t)
\end{align}
with coefficients (see Appendix \ref{AppendixGPEKerrDerivation} for a more rigorous derivation),
\begin{align}
      \omega(t) &\equiv \tilde \omega(t) - \mu(t)\\
    \mu(t) &\equiv \frac{1}{N} i \int_V d^3\! x   \psi^*(\vec x, t )\partial_t \psi(\vec x, t ) \\
    \tilde \omega(t) &\equiv  \frac{1}{N} \frac{\hbar }{2m A^2(t)}\int_V d^3\! x   |\vec{\nabla} \psi(\vec x, t)|^2  \\
    \chi(t) & \equiv \frac{1}{2 N^2} \frac{m}{\hbar} \int_V d^3\! x\, \Phi(\vec x, t)  |\psi(\vec x, t)|^2 \label{Chigeneral}\\
    &= -\frac{1}{2N}  \omega(t)  \label{chiTildeOmegaRelation}\,.
\end{align}
\end{subequations}
The last equality \eqref{chiTildeOmegaRelation} follows from the Gross-Pitaevskii equation (GPE) for $\psi(\vec x,t)$, which is identical in form to the Heisenberg equation \eqref{GPSPEHeisenberg} with all the ``hats'' removed, see Appendix \ref{AppendixGPEKerrDerivation} for a derivation.

Note that the common approach in the ALP literature is to set $\hat \psi(\vec x ,t) \rightarrow \psi (\vec x ,t)$, and thus $\hat a(t) \rightarrow \sqrt{N}$. While this indeed solves \eqref{GPSPKerrE}, it is not strictly a valid solution since it does not satisfy $[\hat a, \hat a^\dagger]=1$. It neglects the quantum noise which undergoes strong squeezing and can therefore grow in time, as we will now show.

The relevant dynamics of $\hat a$ is equivalent to a Kerr-Hamiltonian which is well-known in quantum optics
\cite{TanasKielich1983, YurkeStoler1988}
and in the context of BECs \cite{ParkinsWalls1998}. An initially coherent state $| \alpha=\sqrt{N}\rangle$  evolves unitarily according to the Hamiltonian
$
\hat H = \hbar \omega \hat a^\dagger \hat a + \hbar \chi \hat a^\dagger \hat a^\dagger \hat a  \hat a \,,
$
with time-independent parameters $\omega$ and $\chi$.
The Heisenberg equation of motion $i \hbar \dot {\hat a} = [\hat a, \hat H]$
is equivalent to \eqref{GPSPKerrE} with time-independent coefficients and with solution $ \hat a(t) = e^{- i t (\omega + 2 \hat a^{\dagger}(t_i)\hat a(t_i)\chi) } \hat a(t_i)$.  As we show below, assuming these parameters to be constant is a good approximation for the time-scales we consider.
In the following we set $t_i=0$ to simplify expressions.

\subsection{Quadrature squeezing}
Defining the quadrature operator
$
    \hat X_\theta(t) = \hat a(t) e^{-i \theta}+\hat a^\dagger(t) e^{i \theta} \,,
$
where $\theta$ is the quadrature angle, squeezing is present at $\theta=\theta_-(t) $ if the minimal variance
$
    V_-(t) \equiv \mathrm{Var}(\hat X_{\theta_-(t)}(t)) < 1
$
is below the vacuum level (see Appendix \ref{AppendixSqueezing} for more details).
One effect of the non-linearity is to generate such squeezing, which is still described in the Gaussian approximation and takes place on much shorter timescales than non-Gaussian effects such as generation of superposition states \cite{YurkeStoler1988}. In the limit of large $N$, the minimal variance is given explicitly in Appendix \ref{AppendixKerrModel}, eq.\,\eqref{VminusBajer}.
We define the squeezing timescale as the time on which $V_-(t)$ crosses below $e^{-2}\simeq 0.135$, i.e. the vacuum noise is reduced by a factor 0.135 (or about $9\,\mathrm{dB}$ squeezing). Or in terms of the squeezing parameter
\begin{equation} \label{rDef}
    r(t) \equiv -\tfrac{1}{2} \ln V_-(t) \,,
\end{equation}
the squeezing timescale is $r(t_{\rm sqz})=1$. This yields
\begin{equation} \label{tsqzgeneral}
N |\chi| t_{\rm sqz} \simeq  \tfrac{1}{2} \sinh (1) \simeq 0.6\,.
 \end{equation}
For large $N$, $ V_-(t)$ approaches $0$ very closely, with its mininum corresponding to a maximum of $r(t)$ given
\begin{equation} \label{rminDef}
    r_{\rm max}\simeq \ln\big( 3^{-1/2} 2^{5/6}  N^{1/6} \big) \simeq \ln\big(  N^{1/6} \big)  \,,
\end{equation}
with $r(t_{\rm max})= r_{\rm max}$ and
\begin{equation}\label{tminDef}
   N |\chi| t_{\rm max} \simeq 2^{-5/3}  N^{1/6} \simeq 0.3 N^{1/6} \,.
\end{equation}
The squeezing angle defining the orientation of the squeezed quadrature can be approximated by
\begin{equation} \label{thetamin}
\theta_-(t) \simeq  \mathrm{sgn}(\chi t)\frac{\pi}{4} -\frac{1}{2} \arctan( 2\chi N t) -(\omega + 2\chi N)t \,,
\end{equation}
which for $t> t_{\rm sqz}$ approximately co-rotates with the classical solution
$
    a_c(t)=\alpha e^{-i t( \omega + 2\chi N)} \label{aclassical}
$
of the Kerr model which is a good approximation to the mean $\langle \hat a(t) \rangle$.
The conjugate quadrature with $\theta_+=\theta_- + \pi/2$ is exactly anti-squeezed.
The $N$-dependent phase shift provided by the arc tangent  approximately cancels  for $t> t_{\rm sqz}$ with the remaining offset at $t_{\rm max}$ vanishing as $\mathrm{sgn}(\chi t_{\rm max}) N^{-1/6}$, thus producing an approximately amplitude-squeezed state.\footnote{Eq.\,\eqref{thetamin} corrects previous results, eqs.\,26-28 in \cite{TanasMiranowiczKielich1991}. We also note that our expressions for $r_{\rm max}$ and $t_{\rm max}$ are more accurate than those presented in eq.\,8 of \cite{BajerMiranowiczTanas2002}.}

\subsection{Estimation of the Kerr model parameters}
We can express the parameters \eqref{chiTildeOmegaRelation} that govern the dynamics and the squeezing in terms of known cosmological parameters and the ALP mass.
We  estimate $\tilde \omega(a)$ by treating $\psi(\vec x, t)$ in  perturbation theory (see Appendix \ref{AppendixPert}), or by solving for it numerically using the full GPE. In both cases one approximately finds
\begin{align} \label{vrms2}
   \frac{\hbar }{m}\tilde \omega(A) &= K(A) \simeq \tfrac{1}{2}v^2_{\rm rms} = \tfrac{1}{2} A\,v^2_{\rm rms}(A=1) \,,
\end{align}
where $m K$ is the average ALP kinetic energy, and $v^2_{\rm rms}$ is the density weighted average of the ALP velocity field squared, see
\cite{KoppVattisSkordis2017}, or in the context of cold dark matter (CDM) eq. 106 in \cite{Hertzberg2014}.
Numerical simulations show $v^2_{\rm rms}(A=1)\simeq 10^{-6}$ in the present-day universe.
This is the result of nonlinear structure formation through gravitational instability of $\psi$ from primordial density perturbations.
On scales larger than $\lambdabar_{\rm deB} = \hbar/(m v_{\rm rms})$,  $|\psi(\vec x, t)|^2$ and the density of of cold dark matter are virtually the same \cite{WidrowKaiser1993, SchiveChiuehBroadhurst2014, MoczLancasterFialkovEtal2018, KoppVattisSkordis2017}.
Therefore estimating $K(t)$ can be done with any conventional CDM N-body simulation or CDM perturbation theory.
The squeezing timescales are determined by $\chi$ which is related to the average potential energy $W= \frac{\hbar}{m} N\,\chi $ \cite{KoppVattisSkordis2017}.
Using the Layzer-Irvine equation $\frac{d}{dA}[A (W(A)+K(A))] = - K(A)$, eq. 47 of  \cite{KoppVattisSkordis2017}, we get with $K(A) \propto A$ from \eqref{vrms2}, $W= - \frac{3}{2} K$.
Thus combined
\begin{equation} \label{chiomegarelation}
     \omega(t)  = 3 \tilde \omega(t)\,, \quad
     \chi(t)  = - \frac{3}{2}\frac{\tilde \omega(t)}{N}\,,\quad %
     \mu(t)  = -2 \tilde \omega(t) \,.
\end{equation}
We have therefore tied all parameters appearing in the Kerr oscillator to a simple expression for $\tilde \omega(t)$, eq.\,\eqref{vrms2}, entirely fixed by known cosmological parameters and the ALP mass $m$.

\begin{figure}
        \includegraphics[width=0.49 \textwidth]{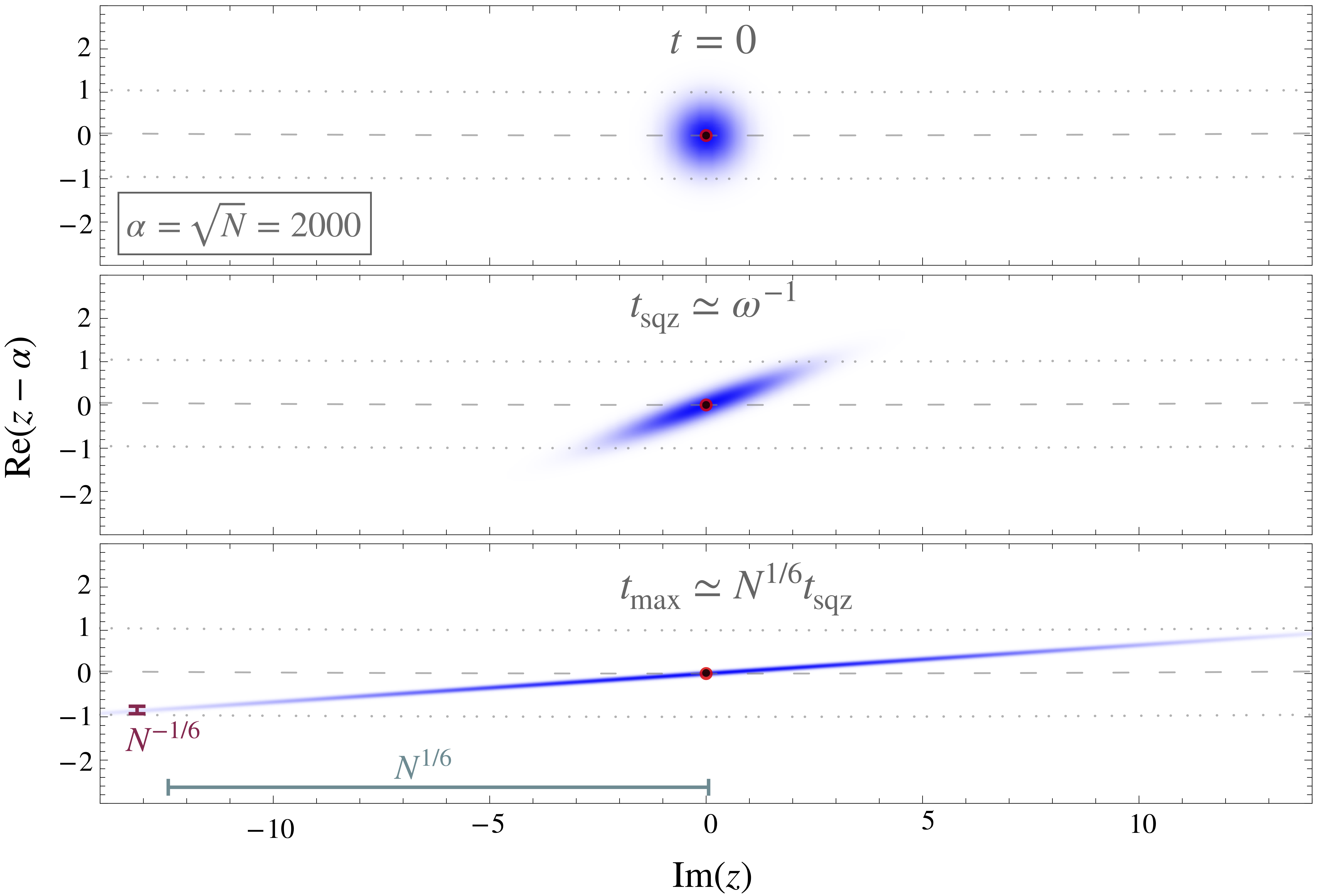}
    \caption{Squeezing generated through unitary time evolution of a dark matter coherent state $| \alpha \rangle $  in the Hartree approximation visualised by the Wigner function \changes{$W(z,t)= (\pi \hbar)^{-1} \langle \alpha | \hat D(z) (-1)^{\hat a^\dagger \hat a} \hat D^\dagger(z) | \alpha \rangle $,  see \cite{CahillGlauber1969}, where $\hat D ( z) = e^{ z \hat a^\dagger -  z^* \hat a}$}. Darker hues of blue correspond to larger values of $W$.  \textit{Top panel}: Initial coherent state. \textit{Middle panel}: Onset of squeezing. \textit{Bottom panel}: Maximal achievable squeezing for $\sqrt{N}=2000$.
The initial time $t=0$ is an arbitrary moment in the late universe (i.e.\,close to today). The red (large) dot indicates the classical solution $a_{\rm cl}(t) = \alpha=2000$, and the black (small) dot the quantum mean field $\langle \hat a(t) \rangle$. The classical solution is time independent since \eqref{chiTildeOmegaRelation} holds for the ALP.}
    \label{fig:squeezingfigure}
\end{figure}

The number $N$ of axions in the volume $V$, with $V^{1/3} \lesssim H^{-1}_{0}$ is approximately
\begin{equation} \label{GPOccupationNumber}
   N \simeq H^{-3}_{0}\frac{\rho_0}{m} =  10^{93} \frac{10^{-5}\,\mathrm{eV}}{m}\,.
\end{equation}
Note that this number deviates from ALP mode occupation numbers used elsewhere in the literature, e.g. $10^{61}$ in \cite{SikivieYang2009} and $10^{26}$ in  \cite{GuthHertzbergPrescod-Weinstein2015} which refer to phase space occupation numbers in the early and late universe, respectively.
Our $N$ is the occupation number of the single GP mode.

\subsection{Onset of squeezing}

We find the squeezing timescale of the ALP in the present-day universe from \eqref{tsqzgeneral}, \eqref{vrms2} and  \eqref{chiomegarelation} to be
\begin{equation} \label{tsqz}
    t_{\rm sqz} \simeq \frac{1}{\tilde \omega} \simeq \frac{\hbar}{m v^2_{\rm rms}} = \frac{\lambdabar_{\rm deB}}{v_{\rm rms}} \stackrel{A=1}{\simeq} \frac{10^{-5}\,\mathrm{eV}}{m}\, 66 \mu\mathrm{s}\,.
\end{equation}
We have thus shown
that squeezing is rapidly generated for an ALP, much faster than the age of the universe $\simeq H_0^{-1} \simeq 10^{18}\,\mathrm{s}$ for all $m > 10^{-22}\,\mathrm{eV}$.
In other words, the single mode treatment of the ALP predicts quantum effects for the full mass range of viable ALP models on cosmologically short timescales. This also justifies our approximation of the Kerr model parameters by the their present-day values, since $|\dot \omega/\omega | \simeq |\dot \chi/\chi | \simeq H$.
\changes{Note that the (quantum) squeezing time scale $\lambdabar_{\rm deB}/v_{\rm rms}$ of the operator $\hat a(t)$  essentially coincides with the (classical) coherence time scale of the GP field $\psi(\vec x, t)$, see for instance \cite{GrahamRajendran2013}.}

For a QCD axion with $m \simeq 10^{-5}\,\mathrm{eV}$, this timescale is about $66 \mathrm{\mu s}$.
One can compare this to other quantum mechanical timescales appearing in the Kerr model, such as the Ehrenfest timescale $t_{\rm Ehr} = N^{1/2} t_{\rm sqz}$, or Schr\"{o}dinger ``cat'' creation time $t_{\rm cat} = N t_{\rm sqz}$ \cite{YurkeStoler1988}.
Both of these timescales far exceed the age of the universe for the ALP mass range.

\begin{table}[t]
{\footnotesize
 \begin{tabularx}{0.48\textwidth}{|l | Y Y Y |}
 \hline
  & Cosmology & Solitonic core & \changes{Milkyway} \\[1ex]
 \hline\hline
 $t_\mathrm{sqz} [\mathrm{\mu s}]$ & $66 \,  \bigl(\frac{10^{-5}\mathrm{eV}}{m}\bigr)$  & $1400 \, \bigl(\frac{10^{-5}\mathrm{eV}}{m}\bigr)$  &  $33\, \bigl( \frac{10^{-5}\mathrm{eV}}{m} \bigr) $  \\[1ex]
  $t_\mathrm{max}$\,[yr] & $3500 \,  (\frac{10^{-5}\mathrm{eV}}{m})^{\frac{7}{6}}$ & $0.5\, \bigl(\frac{10^{-5}\mathrm{eV}}{m}\bigr)^{\frac{4}{3}} $&  $33000 \,  (\frac{10^{-5}\mathrm{eV}}{m})^{\frac{7}{6}}$  \\[1ex]
  $r_{\mathrm{max}}$ & $36+\frac{1}{6}\ln{\bigl(\frac{10^{-5}\mathrm{eV}}{m}\bigr)}$  & $24+\frac{1}{3 }\ln{\bigl(\frac{10^{-5}\mathrm{eV}}{m}\bigr)}$ & $32+\frac{1}{6}\ln{\bigl(\frac{10^{-5}\mathrm{eV}}{m}\bigr)}$\\[1ex]
  $t_{\mathrm{Ehr}}$\,[yr] &  $10^{35}(\frac{10^{-5}\mathrm{eV}}{m})^{\frac{3}{2}}$  &  $10^{21}\bigl(\frac{10^{-5}\mathrm{eV}}{m}\bigr)^2 $ & $10^{32}(\frac{10^{-5}\mathrm{eV}}{m})^{\frac{3}{2}}$
  \\[0.5ex]
 \hline
 \end{tabularx}
 }
 \caption{\small Timescales and size of squeezing due to gravitational self-interactions of the ALP, for various physical scenarios. $t_\mathrm{sqz}$ refers to the timescale on which $9 \mathrm{dB}$ squeezing ($r=1$) is formed, while $t_\mathrm{max}$ is the time for which the maximal squeezing $r_{\mathrm{max}}$ is reached. In comparison, $t_{\mathrm{Ehr}}$ is the Ehrenfest timescale which is often considered as a benchmark for the breakdown of the classical field description. Our results show that quantum effects arise on much shorter timescales.}
 \label{tab:table-name}
\end{table}

\subsection{Maximal squeezing}
The maximum squeezing that can be produced in ALPs is obtained by inserting \eqref{GPOccupationNumber} into  \eqref{rminDef}:
\begin{equation}
    r_{\rm max} \simeq 36 + \frac{1}{6} \ln \frac{10^{-5}\,\mathrm{eV}}{m} \,.
\end{equation}
This immense squeezing is similar in size to the squeezing of inflaton perturbations produced during inflation, which is known to lead to observably large quantum signatures  \cite{GrishchukSidorov1990, AlbrechtFerreiraProkopecEtal1994, PolarskiStarobinsky1996, NelsonRiedel2017}.
The time $t_{\rm max}$, eq.\,\eqref{tminDef}, at which maximal squeezing is reached is
\begin{equation}
    t_{\rm max} = 0.5\, t_{\rm sqz} N^{1/6} \simeq  \Big(\frac{10^{-5}\,\mathrm{eV}}{m} \Big)^{7/6} \, 3500 \,\mathrm{yr} \,.
\end{equation}
Therefore even the extreme squeezing $r_{\rm max}$ would be easily reached in the present universe, justifying our use of constant Kerr model parameters, for $m \geq 10^{-12} \mathrm{eV}$.

Since $\omega + 2N \chi =0$ the squeezing angle evolves slowly between $\theta_- \simeq - \frac{\pi}{4}$ and $\theta_- \simeq - \frac{\pi}{4} N^{-1/6}$ in the interval $ 0 < t \lesssim t_{\rm max }$.
This implies that there is an approximately fixed quadrature for which squeezing remains strongest and grows as $r(t)$ over an extended period $t_{\rm max}$.

As an example we show in Fig.\,\ref{fig:squeezingfigure} the evolution of an initial coherent state with $\alpha=2000$ into a squeezed coherent state using the Wigner representation.
Over time, the quadrature squeezing grows as the quadrature rotates, while the number fluctuation remains constant (see the Appendices \ref{AppendixKerrModel} and  \ref{squeezed_coh_states} for more details).

We also consider two other physical situations with drastically smaller volumes for which the GPE dynamics, or equivalently the Hartree ansatz, may be better justified: \changes{galactic haloes and solitonic cores in dark matter haloes}.
The calculations are analogous to the cosmological case, but without the scale factor in eqs. \eqref{GPSPEHeisenberg}. The details are presented in the Appendices  \ref{AppendixSolitonicCore} and \ref{AppendixHaloscopeMilkyway}, and the results are summarized in Table I. Importantly, the onset of squeezing is independent of $N$, thus similar timescales $t_\mathrm{sqz}$ are found. \changes{In Appendix \ref{AppendixHaloscopeMilkyway} we also show that squeezing of axions contained in a volume that isn't gravitationally bound, such as an axion haloscope. In this scenario the haloscope volume is part of a larger gravitationally bound volume, the galactic halo which is modelled using the Hartree state ansatz adopted in this paper, and exhibit negligible squeezing.}

The simple approximation used here is common in the study of ALPs; it assumes a single GP mode (the Hartree ansatz which assures the validity of the GPE). This GP mode may well approximate some physical situations such as the solitonic core.

\section{Discussion}\label{sec:discussion}
Some important conclusions can be drawn from the results based on this simple model. One is that the quantum effect described here accompanies the dynamics whenever the GP assumption is established through the common Hartree ansatz.
Thus even for large occupancies a single classical field description  that is commonly assumed in ALP cosmology \cite{WidrowKaiser1993, Peebles2000, HuBarkanaGruzinov2000, GuthHertzbergPrescod-Weinstein2015, Hertzberg2016, HuiOstrikerTremaineEtal2017, Hui2021} is incomplete. Our results therefore also highlight the need to scrutinize the range of validity of the GP-ansatz in cosmology, which is at the core of many predictions of ALP behavior such as interference fringes  \cite{SchiveChiuehBroadhurst2014}.
Another conclusion is that the quantum effect we describe would be continuously regenerated on very short timescales even if the ALP state were to collapse onto a coherent state through environmental decoherence.

While a similar non-relativistic QFT description for ALPs was previously used  \cite{Hertzberg2016, ErkenSikivieTamEtal2012, BanikChristophersonSikivieEtal2015, SikivieTodarello2017, ChakrabartyEnomotoHanEtal2018}, the specific quantum effect we isolate here is novel in the context of ALPs and takes place on much shorter timescales than other expected quantum phenomena. This squeezing timescale is related to some previous findings. It matches the ``classical break time'' introduced in \cite{DvaliZell2018}, which characterizes the onset of non-linearities in the classical description. It can also be related to the thermalisation timescale found in \cite{SikivieYang2009, ErkenSikivieTamEtal2012}, where a plane wave mode expansion was used to study thermalisation of macroscopically occupied modes. The relaxation of these modes forms a Bose-Einstein condensate (BEC) with an extension of the entire Hubble patch \cite{ErkenSikivieTamEtal2012}. This thermalisation timescale can be related to our squeezing timescale as $H t_{\rm th}  \simeq \frac{1}{H t_{\rm sqz}}$. Thus mode thermalisation and self-squeezing are efficient at distinct periods with thermalisation happening before self-squeezing.

It is instructive to compare our results to quantum effects in laboratory BECs. Quantum revivals due to the Kerr effect have been demonstrated \cite{GreinerMandelHaenschEtal2002}, as well as spin squeezing of internal states \cite{gross2012spin}. The ground state of a trapped BEC has been shown to be a squeezed displaced state \cite{DunninghamCollettWalls1998}, as found here, see the Appendix \ref{AppendixGroundState}. However, in lab experiments the trapping typically dominates self-interactions and the free-fall time is very short. Thus dynamical generation of self-squeezing as described here has not been observed yet, but related theory for cold atoms has been developed \cite{JohnssonHaine2007,wuster2008quantum,HaineJohnsson2009} and experiments with self-squeezed BECs might become feasible in the near future.

\changesNew{From a purely theoretical perspective, our results show that quantum effects are present in ALPs even in the limit where one typically assumes a purely classical description -- the GP equation from the Hartree ansatz is in fact accompanied by a quantum effect that cannot be avoided. But experimentally, it would be a challenging task to show non-classicality, more formidable than verification of ALPs in the first place. To understand this better, let us consider the verification of squeezing of light which has been developed theoretically and experimentally in the 80's using homodyne detection \cite{YuenChan1983, AbasChanYee1983}, which measures the squeezed quadratures directly but which also requires control of a reference coherent state.   Axions, if they exist, would be very difficult to control due to their feeble interactions such that a homodyne detection scheme and other forms of quantum state tomography, see e.g. \cite{LvovskyRaymer2009}, that require control of a reference quantum state seem unrealisable for the axion. Thus the proposals to detect squeezing in condensed cold atoms mentioned in the previous paragraph, e.g.\,\cite{JohnssonHaine2007,wuster2008quantum,HaineJohnsson2009},  might not be applicable to axions. Another idea is to attempt to transfer the squeezing of the axion field into a squeezing of the haloscope's cavity mode and then infer the squeezing of that electromagnetic mode using the methods for detecting squeezed light.
This approach might not be able to detect the Kerr-squeezing of the Hartree mode because the squeezing of the axion field on scales of the haloscope volume is severely reduced, see Appendix \ref{AppendixHaloscopeMilkyway}.
 Thus, the indirect tests through interferometeric correlation functions as mentioned in the Introduction might be better suited for squeezing verification in axions.
This could be achieved by measuring the intensity correlation function $G_I^{(2)}(t-t')=\langle \hat a^\dagger(t) \hat a^\dagger(t') \hat a(t)  \hat a(t') \rangle$ \cite{Walls1983, Leuchs1986}. At $t=t'$ squeezed states of light can have $ 0< G_I^{(2)}/\langle \hat a^\dagger \hat a \rangle < 1$ so that detected photons are ``anti-bunched'', whereas for a coherent state this is equal to 1 as a consequence of the Poissonian statistics. Any other classical phase space distribution of the electromagnetic radiation field can only exceed this value, which is why squeezed light is called non-classical. In the case of Kerr-squeezing discussed in this paper, however, the squeezing anti-bunching does not occur, see the dashed line in Fig.\,\ref{fig:VarnSqueezedState}. This is related to the fact that the fundamental dynamics preserves the number state. It is thus expected that detecting the squeezing of the axion, given a detection of the axion in the first place, will be a difficult task and will require new approaches or the study of indirect consequences on other fields. }

A multimode treatment will not necessarily inhibit squeezing, which is an inherent consequence of the self-interactions. For instance 
number-squeezing was experimentally observed in a multi-well-trapped BEC \cite{OrzelTuchmanFenselau2001}, \changes{and in the context of scalar field dark matter quadrature squeezing of the GP mode has been observed to persist in a toy model where the Hartree ansatz was replaced by five Fourier modes \cite{EberhardtZamoraKoppEtal2022}. }
Multiple localisation sites, analogs of ALP DM haloes, are known to lead to the emergence of several macroscopically occupied GP-like fields \cite{CederbaumStreltsov2003}. A systematic approach that reveals the necessity for and determines the dynamics of additional mode functions has been established in \cite{AlonStreltsovCederbaum2007, AlonStreltsovCederbaum2008}, and applied to what could be considered analog situations of ALP cosmology  \cite{KatsimigaKoutentakisMistakidisEtal2017, SreedharanChoudhuryMukherjeeEtal2020}.
These latter studies revealed that collision events between BEC solitons composed of attractively interacting atoms can necessitate a dynamical increase of required c-number fields, even if the initial state was accurately described by a single GP field. Large squeezing is also prone to decoherence when additional interactions
\changes{are taken into account \cite{Zurek2001,NelsonRiedel2017}. For example, adding baryons would a priori prevent the pure state Hartree ansatz  \eqref{GPSPEHartree} for the ALP that is used throughout this paper.  Other extensions of relevance will be the addition of unresolved modes of the ALP or even more exotic coupling mechanisms such as time-dilation induced effects \cite{pikovski2015universal}. These will add to the decoherence of the system and likely limit the attainable squeezing.}
But since it is constantly regenerated we expect squeezing to persist at least on some time-scales. The multi-mode treatment\changes{, inclusion of environments} and resulting decoherence will be considered in subsequent work.

\changesNew{Note added: After submission of this article a related work appeared \cite{KussMarsh2021}, in which two-mode squeezing of the axion is discussed. The mechanism and the type of squeezing studied in \cite{KussMarsh2021} is different from that discussed here. In our mechanism squeezing is due to self-gravity, i.e. the nonlinear self-interaction of inhomogeneities of the scalar field in the late universe, whereas in \cite{KussMarsh2021} squeezing of the  inhomogeneous modes of the axion is a linear process due to the strong time-dependence of the expansion of space in the very early inflationary universe \cite{GrishchukSidorov1990, AlbrechtFerreiraProkopecEtal1994, PolarskiStarobinsky1996, NelsonRiedel2017}.} 

\section{Conclusion}
In this paper we have shown quantum squeezing in ALP models of dark matter, using the Hartree approximation which guarantees the validity of the Gross-Pitaevskii equation. Our results highlight the quantum nature of the models, even when they are expected to yield purely classical results. Non-classical squeezed states are continuously formed at very short timescales in different physical scenarios. \changes{On the one hand, our results motivate searches for observable signatures of quantum effects of ALP dark matter}. On the other hand, the results challenge the validity of the simple models that are routinely employed in ALP cosmology.

We thank Tom Abel, Alexander Balatsky, Andrew Eberhardt, Benjo Fraser, David Marsh and Frank Wilczek for helpful discussions. This work was supported by the Swedish Research Council under grant no. 2019-05615, the European Research Council under Grant No. 742104 and by the Branco Weiss Fellowship -- Society in Science.

\def\aj{AJ}
\def\aap{A\&A}
\def\apj{ApJ}
\def\aapr{A\&A Rev.}
\def\apjl{ApJ}
\def\mnras{MNRAS}
\def\araa{ARA\&A}
\def\aj{AJ}
\def\qjras{QJRAS}
\def\physrep{Phys. Rep.}
\def\nat{Nature}
\def\aaps{A\&A Supp.}
\def\apss{Ap\&SS}      
\def\apjs{ApJS}
\def\prd{Phys. Rev. D}
\def\jcap{JCAP}
\def\nar{New Astron. Rev}
\bibliography{references.bib}

\newpage
\appendix
\section{Quadrature variance and squeezing}
\label{AppendixSqueezing}
A quadrature is defined as
\begin{equation}
     \hat X_\theta(t) = \hat a(t) e^{-i \theta}+\hat a^\dagger(t) e^{i \theta} \,,
\end{equation}
and has variance
\begin{align}
    \mathrm{Var}(\hat  X_\theta(t)) & =  1+ 2 (\langle \hat a^\dagger(t) \hat a(t)\rangle - |\langle \hat a(t)\rangle|^2 ) \nonumber \\
    & \quad + e^{-2i \theta} \mathrm{Var}(\hat a(t)) + e^{2i \theta} \mathrm{Var}(\hat a^\dagger(t))
\end{align}
where $\mathrm{Var}(\hat  O) \equiv \langle \hat O^2 \rangle - \langle \hat O \rangle^2 $.
The minimising angle $\theta_-$ is
\begin{equation} \label{gneralthetaminus}
    e^{2 i \theta_-(t)} = - \sqrt{\frac{\mathrm{Var}(\hat a(t))}{\mathrm{Var}(\hat a^\dagger(t))}}
\end{equation}
leading to the minimal variance $V_-(t) \equiv \mathrm{Var}(\hat  X_{\theta_-}(t)) $,
\begin{equation}\label{generalVminus}
    V_-(t) = 1+ 2 (\langle \hat a^\dagger(t) \hat a(t)\rangle - |\langle \hat a(t)\rangle|^2 )  - 2 |\mathrm{Var}(\hat a(t)) |\,.
\end{equation}
Since $\mathrm{Var}(\hat a^\dagger(t)) = (\mathrm{Var}(\hat a(t)))^*$ the only necessary ingredients are $\langle \hat a(t)\rangle$,  $\langle \hat a^2(t) \rangle$ and  $\langle \hat a^\dagger(t) \hat a(t) \rangle$.
\section{Details of the Kerr model}
\label{AppendixKerrModel}

It is convenient to write
$\hat H = \hbar (\omega - \chi) \hat n + \hbar \chi \hat{n}^2$
because the linear and nonlinear parts of $\hat H$ commute in this form so that $U = e^{- i \hat H t/\hbar} = e^{-i (\omega -\chi)t  \hat n } e^{-i \chi t  \hat n^2 }$.
For a coherent state $| \alpha \rangle$ this implies $\langle \hat a^\dagger \hat a \rangle = |\alpha|^2 = N$.
We assume again w.l.o.g that $\alpha = \sqrt{N}$.

\begin{figure}
        \includegraphics[width=0.48 \textwidth]{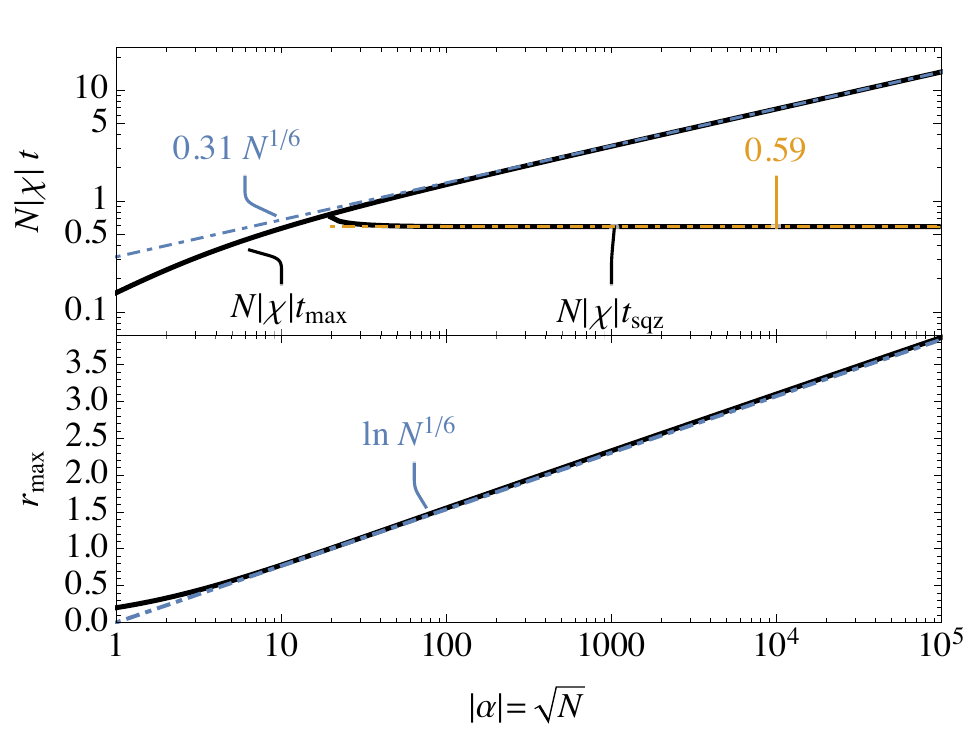}
    \caption{Squeezing timescales $t_{\rm sqz}$ and $t_{\rm max}$ (upper panel) and maximal squeezing (lower panel), and their scaling with $N$,  for the Kerr model. Full lines are exact, dashed lines approximations used in the main text.}
    \label{fig:KerrsqeezingofalphaN}
\end{figure}

To evaluate $V_-$ and $\theta_-$ in the Kerr model, the only additional quantities we need are $\langle \hat a \rangle$ and  $\langle \hat a^2 \rangle$.
We calculate these expectation values using the  result
$e^{\hat A} \hat B  e^{-\hat A} = \sum^\infty_{k=0} \tfrac{1}{k!}[\hat A, \hat B]_k$,
where $[\hat A, \hat B]_k = [\hat A, [\hat A, \hat B]_{k-1}]$,
$[\hat A, \hat B]_{0}= \hat B$.
Assigning  $\hat A = i \hat H t/\hbar$ and $\hat B= \hat a$, or $\hat B = \hat a^2$,  gives $\hat U^{-1} \hat a \hat U$ and $\hat U^{-1} \hat a^2 \hat U$ which can be evaluated using $[\hat a, \hat a^\dagger]=1$, results in
\begin{align}
    \langle \hat a \rangle &=  \langle \alpha| \hat U^{-1} \hat a \hat U | \alpha \rangle \nonumber \\ \label{ameanKerr}
    &= \alpha e^{- i \omega t} e^{-2 N \sin^2(\chi t)} e^{-i N \sin( 2 \chi t)} \\
    \langle \hat a^2 \rangle &=  \langle \alpha| \hat U^{-1} \hat a^2 \hat U | \alpha \rangle  \nonumber \\
    &= \alpha^2 e^{- i 2 (\omega + \chi) t} e^{-2 N \sin^2(2 \chi t)} e^{-i N \sin( 4 \chi t)} \label{a2meanKerr}
\end{align}
Inserting this into \eqref{gneralthetaminus} and \eqref{generalVminus} gives the squeezing angle $\theta_-(t)$ and size $V_-(t)$ for the Kerr oscillator.
The resulting expressions are difficult to analyse further analytically due to their non-algebraic structure.
In \cite{BajerMiranowiczTanas2002} a very accurate algebraic approximation for $V_-(t)$ has been obtained by substituting $N \rightarrow \tau/(\chi t)$ and performing a Taylor expansion in $t$ (with $\tau$ fixed) which gives
\begin{align} \label{VminusBajer}
    V_-(t) &\simeq 1 - 4 \tau s+ 8 \tau^2   + \frac{8 \tau^3(5+12 \tau^2)}{N s} - \frac{16 \tau^4}{N}\\
    s & \equiv \sqrt{1+ 4 \tau ^2} \nonumber \\
    \tau & \equiv N |\chi| t \nonumber \,.
\end{align}
Note that $V_-(t)$ is independent of the sign of $\chi$.
To obtain the squeezing timescale we hold $\tau$ fixed while taking the limit $N \rightarrow \infty$ in \eqref{VminusBajer}. Then we solve $-\tfrac{1}{2} \ln V_-(t_{\rm sqz}) = 1 $ for $t_{\rm sqz}$ to obtain \eqref{tsqz}.
To obtain the time of maximal squeezing we replace $|\chi| t \rightarrow \tilde \tau N^{-5/6}$ in \eqref{VminusBajer}, and while holding $\tilde \tau$ fixed, we expand in $1/N$. The leading term is
\begin{align}
    V_-(\tilde \tau) \simeq  N^{-1/3} \frac{2^9 \tilde \tau^6+1}{2^4 \tilde \tau^2} \,.
\end{align}
Taking a $\tilde \tau$ derivative, we find the minimum at $\tilde \tau_{\rm max} = 2^{-5/3}$, which gives \eqref{tminDef}.
Inserting this back into $V_-(\tilde \tau_{\rm max})$ corresponding to $r_{\rm max} = -\frac{1}{2}\, \ln V_-(\tilde \tau_{\rm max})$  and results in \eqref{tminDef}.
Fig.\,\ref{fig:KerrsqeezingofalphaN} shows $t_{\rm sqz}$, $t_{\rm max}$ and $r_{\rm max}$ as a function of $N$, and compares our approximations (dot-dashed) for \eqref{tsqzgeneral}, \eqref{tminDef} and \eqref{rminDef} to their exact result (full lines).

Finally we calculate $\theta_-$. We proceed in a similar fashion as we did for $V_-$. We insert \eqref{ameanKerr} and \eqref{a2meanKerr} into $\mathrm{Var}(\hat a)/(\mathrm{Var}(\hat a))^*$ and we replace $N \rightarrow T/(\chi t)$. Then we take the log and Taylor expand the result in $\chi t$ with the result
\begin{align}
    \theta_- &\simeq - \frac{i}{4} \ln \Big( \ e^{- i 4 \omega t} \frac{e^{- 8 i  T} (2 T+i)}{2 T -i} \Big) \\
    & =  -\frac{i}{4} \ln \Big( \frac{ 2 T+i}{2 T -i} \Big) -  \omega t - 2  T\,.
\end{align}
This can be simplified into the expression \eqref{thetamin}. The signum function in \eqref{thetamin} arises due a branch cut in the log.

\section{Squeezed coherent states}
\label{squeezed_coh_states}
Displacement operator
\begin{equation} \label{DOperatorDef}
    \hat D ( \beta) = e^{ \beta \hat a^\dagger -  \beta^* \hat a}
\end{equation}
and squeeze operator
\begin{equation} \label{SOperatorDef}
\hat S(\zeta) = e^{\frac{1}{2}(\zeta^* \hat a^2 - \zeta (\hat a^\dagger)^2)}
\end{equation}
have the properties
\begin{align}
    \hat D^\dagger(\beta) \hat a \hat D(\beta) &= \hat a + \beta \\
    \hat S^\dagger(\zeta) \hat a \hat S(\zeta) &= \hat a \cosh(\rho) - \hat a^\dagger e^{2i \phi} \sinh(\rho) \,,
\end{align}
where
\begin{equation}
    \zeta = \rho  e^{2i \phi}
\end{equation}
with $\rho \geq 0$. A squeezed coherent state is defined as
\begin{equation} \label{squeezedcohstatedef}
    | \beta, \zeta \rangle = \hat D ( \beta)  \hat S(\zeta) |0 \rangle \,.
\end{equation}
Evaluating \eqref{generalVminus}, \eqref{gneralthetaminus} we find
\begin{align}
    \langle \hat a \rangle &= \beta \\
    V_- &= e^{-2 \rho} \\
    \theta_- &= \phi \,.
\end{align}
Since we know $\langle \hat a(t) \rangle$, $V_-(t) =: e^{-2 r(t)}$ and $\theta_-(t)$ for the Kerr oscillator, see \eqref{ameanKerr}, \eqref{VminusBajer}, \eqref{thetamin}, we can construct an approximate solution to the Kerr oscillator quantum state in the Schr\"odinger picture as $| \Psi^{\rm SC}(t) \rangle  \simeq | \langle \hat a(t) \rangle , r(t) e^{2 i \theta_-(t)} \rangle$.
In the main text we have approximated $\langle \hat a \rangle \simeq a_c(t)$. This is justified because we consider quantum evolution only up to the time $t_{\rm max} \simeq 0.5 N^{1/6} t_{\rm sqz}$ which is much smaller than the Ehrenfest time $t_{\rm Ehr} = N^{1/2} t_{\rm sqz}$.

The squeezed coherent state approximation breaks down due to non-Gaussianities around the time $t_{\rm max}$.
\begin{figure}
        \includegraphics[width=0.45 \textwidth]{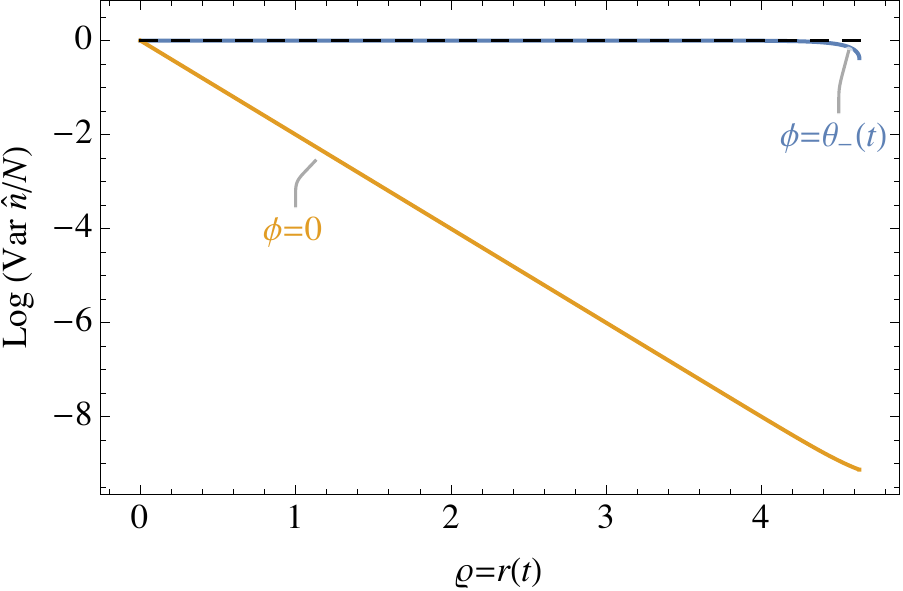}
    \caption{$\mathrm{Var}(\hat n)$ for a squeezed coherent state $| \beta, \rho  e^{2i \phi} \rangle $ with $\beta=\sqrt{N}=10^6$ and $\rho= r(t)$ with $0\leq r(t) \leq r_{\rm max} $ of the Kerr oscillator \eqref{VminusBajer}. Inserting $\phi=0$ corresponds to amplitude squeezing, accompanied by $\hat n$-squeezing, whereas inserting $\phi=\theta_-(t)$ of the Kerr evolution prevents $\hat n$-squeezing. The dashed line shows $\mathrm{Var}(\hat n)$ for a reference Kerr state.}
    \label{fig:VarnSqueezedState}
\end{figure}
In order to verify that a squeezed coherent state is a good approximation to the Kerr-evolved coherent state until $t_{\rm max}$ we evaluate
\begin{equation}
    \mathrm{Var}(\hat n)= \tfrac{1}{4} \big(\cosh(4 \rho)-1\big)  +  N (\cosh(2 \rho) - \cos(2 \phi) \sinh(2 \rho))
\end{equation}
for the squeezed coherent state $| \sqrt{N} , r(t) e^{2 i \phi} \rangle$ with $\phi=\theta_-(t)$ and $\phi=0$ in Fig.\,\ref{fig:VarnSqueezedState}.
We see that inclusion of the squeezing orientation is crucial to match the  value $\mathrm{Var}(\hat n)=N$ of the Kerr-evolved coherent state.
This also explains the ``coincidence'' of $\theta_-(t_{\rm max}) \simeq \tfrac{\pi}{4} N^{-1/6}$ and $V^{1/2}_+ \simeq N^{1/6}$. As can be seen in the lower panel of Fig.\,\ref{fig:squeezingfigure}, this combination of angle and vertical extend of the phase space distribution leads to an order unity range in radial direction that is independent of $N$.


\section{Perturbation theory estimate of $K(A)$}
\label{AppendixPert}
The goal of this Appendix is to show the physical origin, the approximate order of magnitude, as well as the  $A$-scaling of the kinetic energy $K(A)$. This result was used derive a relation \eqref{chiomegarelation} between $\mu$ and $\tilde \omega$ appearing in the Kerr oscillator equation \eqref{GPSPKerrE}.
A more precise estimate of $K(A)$ could be extracted from a cosmological N-body or Schr\"odinger-Poisson simulation if needed.

The mean kinetic  energy $m K(A)$ and the mean potential energy $m W(A)$ of an ALP can be evaluated in perturbation theory. Here we use linear perturbation theory of cold dark matter modelled by a pressureless perfect fluid, with particle density $ n(\vec x, t) $ and velocity field $\vec{u}(\vec x, t)$ satisfying the Euler-Poisson equation.
Since approximately $\tfrac{\hbar^2}{m^2}|\vec{\nabla} \psi(\vec x, t)|^2 = n(\vec x, t) |\vec{u}(\vec x, t)|^2$, see e.g. \cite{WidrowKaiser1993}, \cite{KoppVattisSkordis2017}, the mean kinetic energy per mass is
\begin{align}
    K(A) & = \frac{1}{2 A^2 N}\int_V d^3\!x\ n(\vec x, A) \vec{u}^2(\vec x, A) \\
    &= \frac{1}{2 N}\int_V d^3\!x\ n(\vec x, A) \vec{v}^2(\vec x, A)
\end{align}
where $\vec{v}= \vec{u}/A$ is the peculiar velocity field, and $\vec{u}$ is the canonical velocity field.
In linear perturbation theory the comoving density field can be written as
$n(\vec x, A) = n_0(1 + \delta(\vec x, A))$ with the density perturbation $\delta(\vec x, A)$ related to the velocity field via $\vec{\nabla}\cdot \vec{v} = - A H\delta(\vec x, A)$.
During the assumed matter dominated expansion $\dot A/A= H=H_0 A^{-3/2}$ we thus get to leading order in perturbation theory
\begin{align}
    K(A) &\simeq \frac{1}{2 V}\int_V d^3\!x\  \vec{v}^2(\vec x, A) \\
    &= \frac{A}{2 V}\int_V d^3\!x\ \vec{v}^2(\vec x, A=1)\,.
\end{align}
In Fourier space, taking the limit $V \rightarrow \infty$, this becomes
\begin{align}
    K(A) &\simeq \frac{A}{2 } H_0^2 \int \frac{d^3 k}{(2\pi)^3}  \frac{|\delta_{\vec k}( A=1)|^2}{k^2} \\
    & = \tfrac{1}{2} A\, v^2_{\rm rms}(A=1) \\
    v^2_{\rm rms}(A=1) & := H_0^2  \int \frac{dk}{2\pi^2} P(k,A=1)\,,
\end{align}
where we used the ergodic theorem to replace $|\delta_{\vec k}|^2 \rightarrow P(k=| \vec k|)$, and performed the angular integrals.
For a standard matter power spectrum $P(k)$ this evaluates to $10^{-5}$. The actual value of $v^2_{\rm rms}$ used in the main text is about a ten times smaller due to our negligence of the cosmological constant. Inclusion of a cosmological constant in linear perturbation would give rise to an overall pre-factor $f^2 = 0.27$ and slightly shallower scaling than $A$\changes{, namely $K \propto (A H f)^2$.
Here $f \simeq \Omega_m^{6/11}$ is the growth rate of linear matter density perturbations, $\Omega_m =\Omega_{m,0} A^{-3}H_0^2/H^2$ the fraction of matter and $H = H_0 (\Omega_{m,0} A^{-3} + 1- \Omega_{m,0} )^{1/2}$ the expansion rate.}


\section{Derivation of the coupled GPE and Kerr oscillator} \label{AppendixGPEKerrDerivation}
Throughout this section we work in the Schr\"odinger picture, so that the quantum state $| \Psi(t) \rangle$ of the non-relativistic ALP is time dependent.
For simplicity we discard the cosmological evolution, so that the Hamiltonian is
\begin{multline}
    \hat{\mathcal H} = \int d^3 \!x\, \frac{\hbar^2}{2 m} \nabla \hat \psi^\dagger(\vec x) \nabla \hat\psi(\vec x) - \\ \frac{Gm^2}{2} \int d^3 \!x\,d^3 \!x'\,\frac{\hat\psi^\dagger(\vec x) \hat\psi^\dagger(\vec x') \hat\psi(\vec x)\hat\psi(\vec x')}{|\vec x - \vec x'|} \,.
\end{multline}
$| \Psi(t) \rangle$ satisfies the Schr\"odinger equation
\begin{equation}
     i \hbar \partial_t | \Psi(t) \rangle = \hat{\mathcal{H}} | \Psi(t) \rangle \,.
\end{equation}

The Hartree ansatz for a $n$-particle wave function in the Schr\"odinger picture is
\begin{equation}
\Psi_n(\vec x_1, ..., \vec x_n,t) = \phi(\vec x_1, t) ... \phi(\vec x_n, t)\,.
\end{equation}
Here we introduced the normalised one-particle wave function $\phi(\vec x, t)$.
This $n$-particle state can be written basis-independent as $| n, t \rangle = \frac{1}{\sqrt{n!}} (\hat b^\dagger(t))^n |0 \rangle $, where $\hat b^\dagger(t) \equiv \int d^3 \!x\, \phi(\vec x, t) \hat \psi^\dagger(\vec x)$ and  $\hat \psi^\dagger(\vec x)$ is the time-independent field operator in the Schr\"odinger picture.
A generic (e.g.\,initially coherent) Hartree state $| \Psi(t) \rangle$ then takes the form
\begin{equation}
    | \Psi(t) \rangle = \sum_{n=0}^{\infty} c_n(t) | n, t \rangle \,.
\end{equation}
Our goal is to derive evolution equations for $\phi(\vec x, t)$ and $c_n(t)$.
Varying the action  $S = \int dt \langle \Psi(t) | \hat{\mathcal{H}} - i \hbar \partial_t | \Psi(t) \rangle$ w.r.t.\,$\phi^*(\vec x,t)$ and $c_n^*(t)$ gives
\begin{align}
 i \hbar \partial_t \phi(\vec x,t) & = - \frac{\hbar^2}{2 m} \nabla^2 \phi(\vec x ,t) +  \label{phiEvoVariation1}\\
 & \quad \ + \frac{\overline{ n^2}(t)-\bar n(t)}{\bar n^2(t)} m \Phi(\vec x, t) \phi(\vec x, t)  \nonumber \\
 \nabla^2 \Phi(\vec x, t) & \equiv 4 \pi G m \bar n(t) |\phi(\vec x, t)|^2 \label{phiEvoVariation2}\\[2ex]
    i \hbar \partial_t c_n(t) &= \langle n, t | \hat{\mathcal{H}} - i \hbar \partial_t   | n, t \rangle c_n(t) \\
    \langle n, t | \hbar^{-1}\hat{\mathcal{H}} - i  \partial_t   | n, t \rangle
&= n ( \omega(t) -\chi(t) ) + n^2 \chi (t)\\[2ex]
\bar n(t) &\equiv \sum_n |c_n(t)|^2 n \\
\overline{ n^2}(t) &\equiv \sum_n |c_n(t)|^2 n^2 \\
\omega(t) &\equiv \tilde \omega(t) - \mu (t)\\
\tilde \omega(t) &\equiv  \int d^3 \!x\,  \frac{\hbar}{2 m} |\nabla \phi(\vec x,t)|^2\\
\mu(t) &\equiv -i  \int d^3 \!x\, \phi^*(\vec x, t) \partial_t\phi(\vec x, t) \\
\chi(t) &\equiv \frac{m}{2 \hbar \bar n(t)} \int d^3 \!x\, \Phi(\vec x, t) |\phi(\vec x, t)|^2 \,.
\end{align}
Neglecting the time dependence of $\omega$ and $\chi$ we find
\begin{equation}
    c_n(t) = c_n(t\!=\!0) e^{-it ((\omega-\chi)n + \chi n^2)} \,, \label{KerrSchroedingerPic}
\end{equation}
which is the solution of the Schr\"odinger equation for a Kerr oscillator.
For an initial coherent state $| \Psi(t=0) \rangle = | \alpha \rangle$ with $\alpha = \sqrt{N}$ we have $c_n(t\!=\!0) = e^{-N/2} \frac{N^{n/2}}{\sqrt{n!}}$ so that $\bar n(t) = N$ and $\overline{n^2}(t) = N^2+N$.
The $\phi$ evolution equation, \eqref{phiEvoVariation1} and \eqref{phiEvoVariation2}, then simplifies to
\begin{subequations}
\begin{align}
 i \hbar \partial_t \phi(\vec x,t) & = - \frac{\hbar^2}{2 m} \nabla^2 \phi(\vec x ,t)  +  m \Phi(\vec x, t) \phi(\vec x, t) \\
 \nabla^2 \Phi(\vec x, t) & \equiv 4 \pi G m N |\phi(\vec x, t)|^2
 \end{align} \label{GPESchroedingerPic}
 \end{subequations}
which is the GPE used in the main text, after re-defining $\psi \equiv \phi \sqrt{N}$ and restoring the scale factor $A(t)$.
 Equations \eqref{KerrSchroedingerPic}, \eqref{GPESchroedingerPic} then establish the validity of the constraints $\int d^3 \!x\, |\phi(\vec x, t)|^2 =1 $ and $\sum_n |c_n(t)|^2=1$, which we could have enforced via Lagrange multipliers \cite{AlonStreltsovCederbaum2007}.
Noticing that $\langle n, t | \hat{\mathcal{H}} - i \hbar \partial_t   | n, t \rangle = \langle n | \hat H | n \rangle$, with $| n \rangle = | n , t\!=\!0\rangle$ and $\hat H = \hbar (\omega - \chi) \hat n + \hbar \chi \hat n^2$, we verify \eqref{GPSPKerrE} for $\hat a(t) = e^{i \hat H t/\hbar} \hat b(t=0) e^{-i \hat H t/\hbar}$ in the Heisenberg picture.

In the main text we have worked in the Heisenberg picture where $\hat \psi(\vec x,t) = \phi(\vec x, t) \hat a(t)$. As a check let us compare the result of the mean field $\langle \alpha | \hat \psi(\vec x,t) | \alpha \rangle = \phi(\vec x, t) \langle \alpha | \hat a(t) | \alpha \rangle$ in the Heisenberg picture to that in
Schr\"odinger picture $\langle \Psi(t) | \hat \psi(\vec x) | \Psi(t) \rangle$.
For this we assume $\hat \psi(\vec x) = \phi(\vec x, t) \hat b(t)$.
This seems somewhat odd since $\hat \psi(\vec x)$ should be time-independent, but is not. This is merely an artifact of restricting the field operator to contain only the single GP mode, and does not pose a problem.
We thus have $\langle \Psi(t) | \hat \psi(\vec x) | \Psi(t) \rangle=\phi(\vec x, t) \langle \Psi(t) | \hat b(t) | \Psi(t) \rangle$.
Let us focus on $\langle \Psi(t) | \hat b(t) | \Psi(t) \rangle$ and switch to the ``Kerr picture''  that we define through  $| \Psi_{\rm K}(t) \rangle \equiv \sum_n c_n(t) |n, t\!=\!0\rangle$.
Clearly, $\langle \Psi(t) | \hat b(t) | \Psi(t) \rangle = \langle \Psi_{\rm K}(t) | \hat b(t\!=\!0) | \Psi_{\rm K}(t) \rangle$.
This equals to $\langle \Psi_{\rm K}(t\!=\!0) | e^{i \hat H t/\hbar}\hat b(t\!=\!0)  e^{-i \hat H t/\hbar} | \Psi_{\rm K}(t\!=\!0) \rangle = \langle \alpha |\hat a(t) | \alpha \rangle$, and we have established the equivalence to the Heisenberg picture calculation.
Similar calculations establish the equivalence of all correlators and thus in particular the squeezing.

In summary, using a variational ansatz for a quantum state in the Hartree approximation we have derived the coupled GP and Kerr oscillator equations for an initial coherent state of a non-relativistic ALP field.

\section{Squeezed state approximation}
\label{AppendixKerrpicture}
In the ``Kerr picture'' where $\hat \psi_{\rm K}(\vec x, t)=\psi(\vec x, t) \tfrac{\hat a(t_i)}{\sqrt{N}}$ and $| \Psi_{\rm K} (t) \rangle = e^{-i \hat H t/\hbar} |\alpha \rangle$, the quantum state can be approximated by a squeezed coherent state
\begin{equation}
    |\Psi^{\rm SC}_{\rm K}(t) \rangle =  \hat D\big(a_c(t) \big) \hat S\Big(r(t) e^{2 i \theta_-(t)}\Big) |0 \rangle \,,\label{ApproxKerrState}
\end{equation}
where $\hat D ( \beta) = e^{ \beta \hat a^\dagger(t_i) -  \beta^* \hat a(t_i)}$, and $\hat S(\zeta) = e^{\frac{1}{2}(\zeta^* \hat a^2(t_i) - \zeta (\hat a^\dagger(t_i))^2)}$, are the displacement and squeezing operators, respectively \cite{Walls1983}. Properties of this state are shown in Fig.\ref{fig:squeezingfigure} and Fig.\,\ref{fig:VarnSqueezedState} (the curve labeled by $\theta_-$).

A ``classical'' wave function $|\Psi^{\rm cl}_{\rm K}(t) \rangle$ would then be a state with   $\hat a(t_i) |\Psi^{\rm cl}_{\rm K}(t) \rangle = a_c(t) |\Psi^{\rm cl}(t) \rangle $ at all times, which is only true for a coherent state $|\Psi^{\rm cl}_{\rm K}(t) \rangle = \hat D(a_c(t)) | 0 \rangle$.
The fidelity $|\langle \Psi^{\rm cl}_{\rm K}(t)| \Psi^{\rm SC}_{\rm K}(t) \rangle|^2 = |\langle 0| \hat S\big(r(t) e^{2 i \theta_-(t)} \big) |0 \rangle|^2 $  decays at the squeezing time  $t_{\rm sqz}$ indicating deviation from the classical approximation.

\section{Ground state in a single-GP-mode approximation}
\label{AppendixGroundState}
If thermalisation as described in \cite{ErkenSikivieTamEtal2012} indeed keeps the axion at its momentary ground state $|\Psi_{0}(t) \rangle $, then it might be natural to expect that this ground state is related to our squeezed state.
\cite{ErkenSikivieTamEtal2012} does not explore the nature of this ground state.
However, the ground state of a self-interacting BEC in the single mode approximation is discussed in  \cite{DunninghamCollettWalls1998, ParkinsWalls1998}.
To leading order in a perturbative solution $\hat a=  \alpha + \delta \hat a$, this ground state is found to be a squeezed coherent state
\begin{equation}
     |\Psi_{0}(t)\rangle \simeq e^{- i \mu_0 t} |\Psi(t_{\rm max}) \rangle \,,
\end{equation}
 matching our approximate solution \eqref{ApproxKerrState} at $t_{\rm max}$ apart from some irrelevant global phase.
Squeezing in the Kerr model reaches a maximum only because the quantum state ceases to be well approximated by a Gaussian after $t_{\rm max}$.
Interestingly, the relation to the BEC ground state might persist even when non-Gaussianities are included:  it was found in \cite{ DunninghamCollettWalls1998,ParkinsWalls1998} that the ground state Wigner function of the Kerr Hamiltonian is bent into a ``banana'' producing a so-called number-squeezed state.

\section{Solitonic cores}
\label{AppendixSolitonicCore}

For the solitonic core case, the relevant parameter to compute in order to obtain the squeezing timescale is\\ \begin{align} \label{chi}
    N\chi & \equiv \frac{1}{2 N} \frac{m}{\hbar} \int_V d^3\! x\, \Phi(\vec x, t)  |\psi(\vec x, t)|^2 \,,
\end{align}\\
where we have assumed that $\chi$ is constant in time for the timescales we are interested. Assuming a spherically symmetric dark matter halo and making use of the radial core density profile in \cite{SchiveChiuehBroadhurst2014}, it is straightforward to find the Newtonian potential by solving the Poisson equation. Inserting \eqref{chi} into \eqref{tsqzgeneral}, we get
\begin{equation}
   t_{\mathrm{sqz}}\simeq2\times 10^{8}\biggl(\frac{m}{10^{-22}\mathrm{eV}}\biggr)\biggl(\frac{r_c}{\mathrm{kpc}}\biggr)^2\mathrm{yr} \,,
\end{equation}
where $r_{\mathrm{c}}$ is the solitonic core radius. It is convenient to express the squeezing timescale in terms of the host halo mass $M_h$. To achieve that, we make use of the relation between the core radius $r_c$ and the core mass $M_c$ \cite{SchiveChiuehBroadhurst2014}
\begin{equation}
    \frac{r_c}{\mathrm{kpc}}=\frac{5.5\times10^7}{\bigl(\frac{m}{10^{-22}\mathrm{eV}}\bigr)^2}\frac{M_{\odot}}{M_c}
\end{equation}
and then express the core mass in terms of the host halo mass $M_h$ \cite{PhysRevLett.113.261302}
\begin{equation}
    M_c=\frac{1}{4\sqrt{\alpha}}\biggl(\frac{\zeta{(z)}}{\zeta{(0)}}\biggr)^{\frac{1}{6}}\frac{\bigr(4.4\times 10^7\bigl)^{\frac{2}{3}}}{\frac{m}{10^{-22}\mathrm{eV}}}M_{\odot}^{\frac{2}{3}}M_{\mathrm{h}}^{\frac{1}{3}} \,,
\end{equation}
where $z=A^{-1}-1$.
For further details, the reader is referred to \cite{PhysRevLett.113.261302}.
The squeezing timescale as a function of the host halo mass is given by
\begin{equation}
    t_{\mathrm{sqz}}=7\times 10^{14} \biggl(\frac{10^{-22}\mathrm{eV}}{m}\biggr)\biggl(\frac{M_{\odot}}{M_{\mathrm{h}}}\biggr)^{\frac{2}{3}}\,\mathrm{yr}\,.
\end{equation}
For a dark matter halo of mass $M_{\mathrm{h}}=2\times 10^{12}M_{\odot}$ we find
\begin{equation}
    t_{\mathrm{sqz}}\simeq1400 \biggl(\frac{10^{-5}\mathrm{eV}}{m}\biggr)\mathrm{\mu s}
\end{equation}
The number of axions inside the solitonic core is
\begin{equation}
N=\frac{M_c}{m}\simeq 3\times 10^{62}\biggl(\frac{10^{-5}\mathrm{eV}}{m}\biggr)^2
\end{equation}
Having obtained the squeezing timescale and the total number of axions inside a solitonic core, it is straightforward to compute $t_{\mathrm{max}}$, $r_{\mathrm{max}}$ and $t_{\mathrm{Ehr}}$. The values can be found in the main text (Table I).

\section{\changes{Axion haloscope and Milkyway}}
\label{AppendixHaloscopeMilkyway}
The Hartree, or single mode, ansatz can only be applied to sufficiently isolated systems. In the present context of a purely dark matter dominated universe this means that the volume to which we can apply the Hartree ansatz must be to a good approximation self-gravitating. This was the case for the entire universe and the solitonic core. The axions within a haloscope volume are not self-gravitating, in the sense that their motion is dominated by the external galactic gravitational potential and not their self-gravity. To remain consistent with the Hartree ansatz, the premise of this work, we assume that the whole galactic halo -- plausibly the smallest  self-gravitating volume containing the haloscope -- is described by a Hartree state under time evolution and then focus on the squeezing within a subvolume comprising the haloscope as the observable. This provides an estimate of the squeezing timescale for axions within a haloscope.

The entire galactic halo contains $N$ axions and the assumed Hartree ansatz implies that the only relevant operator is
\begin{equation}
\hat a = \int d^3 x \,\phi^*(\vec x, t) \hat \psi(\vec x, t)\,,
\end{equation}
 where $\phi(\vec x, t) = \psi(\vec x, t)/\sqrt{N}$ is the 1-particle wave function of a galactic axion.
We now decompose $\phi(\vec x, t)$ into two orthogonal functions $\phi_h := \frac{\sqrt{N}}{\sqrt{N_h}} \,\phi(\vec x, t) \theta_h(\vec x)$ and $\phi_{\bar h} :=  \frac{\sqrt{N}}{\sqrt{N-N_h}} \, \phi(\vec x, t) (1-\theta_{h}(\vec x))$, where $\theta_{h}(\vec x)$ equals 1 within the haloscope and vanishes outside of it, and $N_h$ is the number of axions in the haloscope. We then have
\begin{equation}
\hat a  = \frac{\sqrt{N_h}}{\sqrt{N}} \hat b + \frac{\sqrt{N-N_h}}{\sqrt{N}} \hat c
\end{equation}
with $[\hat a, \hat a^\dagger]=[\hat b, \hat b^\dagger]=[\hat c, \hat c^\dagger]=1$, $[\hat b, \hat c^\dagger] = [\hat c, \hat b^\dagger] =0$ and $[\hat b, \hat a^\dagger] = \frac{\sqrt{N_h}}{\sqrt{N}}$, and
\begin{align}
\hat b =& \int d^3 x \,\phi_{h}^*(\vec x, t) \hat \psi(\vec x, t) \\
\hat c =& \int d^3 x \, \phi_{\bar h}^*(\vec x, t) \hat \psi(\vec x, t) \,.
\end{align}
A similar decomposition has been employed in \cite{Simon2002, Lee2015}  to investigate the entanglement between a subvolume and the remaining part of a BEC.
The squeezing of the haloscope mode $\hat b$ is given by
\begin{equation}\label{bVminus}
    V^h_-(t) = 1+ 2 (\langle \hat b^\dagger(t) \hat b(t)\rangle - |\langle \hat b(t)\rangle|^2 )  - 2 |\mathrm{Var}(\hat b(t)) |\,,
\end{equation}
where we assume as before that the initial quantum state (the quantum state in the Heisenberg picture) of the galactic halo is in an $\hat a(t_i)$-mode squeezed coherent state $|\alpha \rangle$. We have shown that unitary evolution can be approximated by  $\hat U(t) = \hat D(\sqrt{N}) \hat S( r(t) e^{2 i \theta_-(t)}) \hat D^\dagger(\sqrt{N}) $,  see eqs. \eqref{DOperatorDef}, \eqref{SOperatorDef} and \eqref{ApproxKerrState}.
Using the commutation relation between $\hat a$ and $\hat b$ operators we find using a calculation similar to that presented in Appendix \ref{AppendixKerrModel}
\begin{align}
    \hat D^\dagger(\alpha) \hat b \hat D(\alpha) &= \hat b +  \frac{\sqrt{N_h}}{\sqrt{N}} \alpha \\
    \hat S^\dagger(\zeta) \hat b \hat S(\zeta) &=   \frac{\sqrt{N_h}}{\sqrt{N}} \big(\hat a \cosh( r) - \hat a^\dagger e^{2i \theta_-} \sinh( r)\big) \,,
\end{align}
with   $\alpha=\sqrt{N}$ and $\zeta= r e^{2 i \theta_-}$.
Using these results \eqref{bVminus} can be simplified most easily in the ``Kerr picture'', see Appendix \ref{AppendixKerrpicture}, to give
\begin{equation}\label{bVminusRes}
    V^h_-(t) = 1+   \frac{N_h}{N}\Big(V_-(t) -1\Big) \,,
\end{equation}
where  $V_-(t) = e^{-2 r(t)}$ is the minimal variance of  $\hat a$. Note that although this result looks quite intuitive it is nontrivial since the $\hat a$-squeezed coherent state is not a product state of a $\hat b$-squeezed coherent state and a  $\hat c$-squeezed coherent state, such that the haloscope subvolume is entangled with the rest of the halo.

This means that at $t_{\rm sqz}$, defined by $V_-(t_{\rm sqz}) = e^{-2}$, when the galactic $\hat a$ mode gets significantly squeezed, the $\hat b$ mode squeezing is
\begin{equation}\label{bVminusRes}
    V^h_-(t_{\rm sqz}) = 1+   \frac{N_h}{N}(e^{-2} -1) \,,
\end{equation}
or assuming $N_h/N \ll 1$
\begin{equation}\label{bsqz}
    r^h(t_{\rm sqz}) \simeq 0.43 \frac{N_h}{N} \ll r(t_{\rm sqz}) = 1\,.
\end{equation}
Thus, the squeezing of a subvolume ($\hat b$ mode) is significantly reduced compared to squeezing of the full volume ($\hat a$ mode),  and bounded by $N_h/N$, the ratio  of the number of axions in the subvolume and the total number of axions.
Similarly, the maximum squeezing of the $\hat b$ mode is drastically reduced. Since $r_{\rm max} < \infty$, one finds
 \begin{equation}\label{bmax}
    r^h(t_{\rm max}) \lesssim 0.5 \frac{N_h}{N}  \ll r_{\rm max}\,.
\end{equation}
For the galactic halo we assume a stationary Navarro–Frenk–White (NFW) density profile \cite{1996ApJ...462..563N}
\begin{equation}
\label{NFW}
    \rho(r)=\frac{\rho_0}{\frac{r}{R_s}\biggl(1+\frac{r}{R_s}\biggr)^2}
\end{equation}
to calculate $\chi$. The parameters $\rho_0$ and $R_s$ vary from halo to halo. Solving the Poisson equation, the gravitational potential is
\begin{equation}
    \Phi(r)=-\frac{4\pi G\rho_0R_s^2}{\frac{r}{R_s}}\ln\Bigl(1+\frac{r}{R_s}\Bigr)
\end{equation}
so that
\begin{equation}
 N \chi =  \frac{1}{2 } \frac{1}{N \hbar} \int_V d^3\! x\, \Phi(\vec x)  \rho(\vec x) \,.
 \end{equation}
 A sufficiently accurate approximation to the total gravitational energy is given by
 \begin{equation}
 N \chi \simeq  \frac{1}{2 } \frac{M_{\rm MW}}{N \hbar} \bar \Phi \simeq  -\frac{1}{2 } \frac{m}{ \hbar} 10^{-6}
 \end{equation}
where $ \bar \Phi \simeq  10^{-6} $ is the average Galactic potential and $M_{\rm MW}\simeq 10^{12} M_\odot$ the Galactic mass.
Substitution of $N\chi$ into \eqref{tsqzgeneral} gives
\begin{equation}
\label{sqzhalo}
    t_{\mathrm{sqz}}\simeq 30 \biggl(\frac{10^{-5}\mathrm{eV}}{m}\biggr)\, \mathrm{\mu} \mathrm{s} \,.
\end{equation}
To evaluate the timescale of maximal squeezing (and its magnitude expressed by $r_{\mathrm{max}}$) we need
\begin{equation}
N=\frac{M_{\rm MW}}{m}= 10^{83} \biggl(\frac{10^{-5}\mathrm{eV}}{m}\biggr) \,.
\end{equation}
Furthermore, we need to find the approximate number of axions in the axion haloscope volume \cite{PhysRevLett.104.041301}
\begin{equation}
N_h=\frac{M}{m}=\frac{\rho(r=8\mathrm{kpc})V_{h}}{m}\simeq6\times 10^{18}\biggl(\frac{10^{-5}\mathrm{eV}}{m}\biggr) \,.
\end{equation}
Inserting this into \eqref{rminDef} and \eqref{tminDef} we get
\begin{equation}
    r_{\mathrm{max}}\simeq \ln( N^{1/6})\simeq 31.8+\frac{1}{6}\ln\biggl(\frac{10^{-5}\mathrm{eV}}{m}\biggr)
\end{equation}
and
\begin{equation}
    t_{\mathrm{max}}\simeq 0.5 t_{\mathrm{sqz}} N^{1/6} \simeq 33000  \biggl(\frac{10^{-5}\mathrm{eV}}{m}\biggr)^{\frac{7}{6}} \mathrm{ yr}\,.
\end{equation}\\
For comparison, we find that the Ehrenfest timescale is approximately $t_{\mathrm{Ehr}}=\sqrt{N}t_{\mathrm{sqz}}\simeq 10^{32} (\frac{10^{-5}\mathrm{eV}}{m})^{\frac{3}{2}}\mathrm{ yr}$.

Finally, let us evaluate the numerical value of the squeezing of the haloscope mode at $t_{\rm sqz}$ and $t_{\rm max}$.
We find from \eqref{bsqz} and \eqref{bmax} that
\begin{equation}\label{bsqzActual}
    r^h(t_{\rm sqz}) \simeq r^h(t_{\rm max})  \lesssim  0.5 \frac{N_h}{N} = 10^{-65}\,.
\end{equation}
\end{document}